\documentstyle[aps,tighten]{revtex}
\begin{document}
\input{epsf}

\title{Quantum Inequalities for the Electromagnetic Field}

\author{Michael J. Pfenning\footnote{email: mitchel@physics.uoguelph.ca}}

\address{Department of Physics, University of Guelph, Guelph,
         Ontario, N1G 2W1, Canada}

\date{\today}

\maketitle

\begin{abstract}
A quantum inequality for the quantized electromagnetic field is developed
for observers in static curved spacetimes.  The quantum inequality derived
is a generalized expression given by a mode function expansion of the
four-vector potential, and the sampling function used to weight
the energy integrals is left arbitrary up to the constraints that it be a
positive, continuous function of unit area and that it decays at infinity.
Examples of the quantum inequality are developed for Minkowski spacetime,
Rindler spacetime and the Einstein closed universe. 

\vspace*{5pt}
\begin{center}
{\sl Dedicated to the memory of Dr. George Leibbrandt.}
\end{center}
\end{abstract}

\pacs{04.62.+v, 03.70.+k, 11.10.-z, 04.60.-m}

\section{Introduction}

Nearly four decades ago, it was shown by Epstein, Glaser and Jaffe
\cite{Epstein} that a positive definite energy density was incompatible
with the usual postulates of a quantized field theory.  Worse yet, it 
appears that the energy density is not even bounded from below. Thus, 
all standard quantized field theories are capable of violating all
the pointwise and averaged energy conditions in general relativity.
However, this does not mean that the energy density can remain negative
for an arbitrarily long period of time.   Over the last decade, new
forms of energy conditions involving various temporal and spatial
averagings have been developed
\cite{Ford91,Klinkhammer,Wald91,F&Ro95,Yurtsever,F&Ro97,Pfen97a,F&P&R97,Flan97,Song97,Pfen98a,Pfen98b,Fe&E98,Fe&T99,F&Ro99,Helf99,Fe&T00,Fews00,Voll00,Fe&V01}.
One such example is 
the quantum inequality, which is the weighted temporal average of the
energy density along the worldline of an observer.  Derived directly from
quantum field theory, these inequalities limit the magnitude and temporal
duration of existence of negative energy densities.  
The quantum inequalities say that if an observer tries to make a
measurement of the energy density for some characteristic sampling time
$\tau_0$, then the maximal negative energy that he might ever measure is
bounded below by an inverse power of the characteristic sampling time. 
Given an observer's four-velocity $u^\mu$ and a sampling (weighting)
function $f(\tau)$  of characteristic width $\tau_0$, then the quantum
inequality is given by
\begin{equation}
\tilde\rho \equiv \int_{-\infty}^{\infty} \langle T_{\mu\nu}(\tau)
\rangle_{Ren.}\,u^\mu(\tau) u^\nu(\tau) \,f(\tau) d\tau \geq 
-{\alpha \over \tau_0^n} S(\tau_0)+\rho_{vacuum}.
\end{equation}
Here $\alpha$ is a dimensionless constant of order unity and $n$ is the
dimension of the spacetime.   For a massless field in Minkowski spacetime,
the function $S(\tau_0)$ is equal to one and the vacuum energy density 
vanishes.  For massive fields and/or curved spacetimes, $S(\tau_0)$
represents the modification of the quantum inequality away from its massless,
flat space functional form.  It has the generic behavior that it is
approximately one for small $\tau_0$, and in most spacetimes it typically
decays for longer characteristic sampling times. However there are some
known exceptions, such as four-dimensional de~Sitter and Rindler spacetimes,
where the function $S$ only grows only as fast as $\tau_0^2$.

The quantum inequalities were first derived by Ford \cite{Ford91} to
constrain negative energy fluxes for the quantized, massless,
minimally-coupled scalar field in Minkowski spacetime.  These results
were then expanded to the energy density of the massive scalar field
in Minkowski space \cite{F&Ro95,F&Ro97} and in static curved spacetimes
\cite{Pfen97a,Pfen98a,Pfen98b}.  In all of the these cases, a Lorentzian
sampling function,
\begin{equation}
f(\tau) = {\tau_0 \over \pi}{1\over\tau^2+\tau_0^2},
\label{eq:lorentzian}
\end{equation}
was used to simplify the calculations.  However, Flanagan \cite{Flan97} 
showed it was possible to derive optimum quantum inequalities for the
massless scalar field in two dimensions for an arbitrary, smooth positive
choice of the sampling function.  This was followed by the work of Fewster
and colleagues \cite{Fe&E98,Fe&T99,Fews00} who have established the quantum
inequality for the minimally coupled scalar field in static curved spacetimes
of any dimension with an arbitrary, smooth positive sampling function. 

Although much of the previous work has been for the scalar field, work
is now progressing for higher spin fields.  Vollick has shown that an
optimum quantum inequality can be derived for the Dirac field in two 
spacetime dimensions \cite{Voll00} for an arbitrary sampling function
using the conformal properties of the field theory. More recently,
Fewster and Verch have established ``quantum weak energy inequalities''
for the Dirac and Majorana fields of nonzero mass in four-dimensional
globally hyperbolic spacetimes \cite{Fe&V01}.  Making use of microlocal
analysis techniques, Fewster and collaborators \cite{Fews00,Fe&V01} have
vastly extended the applicability of the quantum inequalities to arbitrary
globally hyperbolic spacetimes.

The first quantum inequality for the electromagnetic field was derived by
Ford and Roman \cite{F&Ro97} for a Lorentzian sampling function in flat 
spacetime.  This was immediately generalized to curved static spacetimes
by the author \cite{Pfen98b}, although both of these calculations relied
on the specific choice of the Lorentzian sampling function.  In addition,
the proof in both cases was mathematically long and at some times quite
complicated, particularly in some of the lemmas required.

In this paper, we will show that it is possible to derive a generalized
quantum inequality for the quantized electromagnetic field in static
curved spacetimes with a length element of the form
\begin{equation}
{ds}^2 = -|g_{00}({\bf x})|{dt}^2 + g_{ij}({\bf x})dx^i dx^j.
\label{eq:metric}
\end{equation}
The proof presented here is greatly simplified, in large part due to
generalization of a more direct positivity lemma originally developed
by Fewster and colleagues\cite{Fe&E98,Fe&T99}.  In addition, the electromagnetic
field quantum inequality is proven for an arbitrary choice of sampling
function so long as it be a positive, continuous function of unit area
that decays at infinity.  The end result of our calculations is the
quantum inequality written as a mode function expansion,
\begin{eqnarray}
\tilde\rho &\geq&  -{1\over 2\pi} \int_0^\infty d\nu
\sum_\lambda \int d^3{\bf k} \left| \widehat{f^{1/2}}[ \nu + \omega
({\bf k})] \right|^2 \left[{1\over |g_{00}|} 
\overline{E_i(\lambda,{\bf k};{\bf x})}\; g^{ij} \; E_j(\lambda,{\bf k};{\bf 
x})
\right.\nonumber\\
&&\qquad\qquad+\left.\left|{g_{00}\over g}\right| \overline{B_i(\lambda,{\bf 
k};
{\bf x})}\; (g^{ij})^{-1} \; B_j(\lambda,{\bf k};{\bf x})\right]
+\rho_{\rm vacuum},
\label{eq:EM_Quant_Ineq}
\end{eqnarray}
where $E_j(\lambda,{\bf k};{\bf x})$ and $B_j(\lambda,{\bf k};{\bf x})$
are the modes for the electric and magnetic components of the field-strength
tensor,
\begin{equation}
\widehat{f^{1/2}}(\omega) = \int_{-\infty}^\infty  f^{1/2}(t)
\,e^{-i\omega t}\,dt,
\end{equation}
is the Fourier transform of the square root of the sampling function
and the summation over $\lambda$ and integration over $d^3{\bf k}$ is
over all possible polarizations and momentum eigenstates, respectively.
As was the case for the scalar field, the electromagnetic field quantum
inequality (\ref{eq:EM_Quant_Ineq}) tells us how much negative energy an
observer may measure relative to the vacuum energy of the electromagnetic
field.

In Section~\ref{sec:QED_basics} we will discuss the canonical quantization
of the electromagnetic field in curved space and elucidate the particle
state structure and the form of the stress-tensor.  In particular, two
different forms of quantization will be discussed: direct quantization
in the classical Coulomb gauge and the more elegant Gupta-Bleuler form
of quantization. In Section~\ref{sec:positivity} we develop the positivity 
lemma for generic inner-products of vector fields, which is a generalization
of work developed by Fewster and colleagues\cite{Fe&E98,Fe&T99} for the scalar
field.  In Section~\ref{sec:the_QI} we lay out the remainder of the proof
of the quantum inequality, finally arriving at the expression above.
Lastly, in Section~\ref{sec:examples} we will look at the resulting
quantum inequalities for Minkowski spacetime, Rindler spacetime and the
Einstein closed universe.

We will follow the the convention of Wald \cite{Wald} where the
signature of the metric is $(-,+,+,+)$.  Greek indices are summed 
over (0,1,2,3) while Latin indices denote the spatial components
(1,2,3).  However, the letter $\lambda$ has been singled out as the
polarization state label, and depending on the context, can represent
either the two physical polarization states 1 and 2, or the full
set of polarization states 0, 1, 2 and 3 in the Gupta-Bleuler formalism
which includes the scalar and axial photon polarization states.
Also, the complex conjugate of $f$, will be denoted by $\overline{f}$.
Units of $\hbar = c = G = 1$ will be used throughout.
%%%%%%%%%%%%%%%%%%%%%%%%%%%%%%%%%%%%%%%%%%%%%%%%%%%%%%%%%%%%%%%%%%%%%%%%%%%%%%%%%%
\section{Electromagnetic Field in Static Curved Spacetimes}
\label{sec:QED_basics}

We begin our discussion of the electromagnetic field by defining the
classical Maxwell action for a source free field in curved space,
\begin{equation}
S^{Maxwell} = - {1\over 4} \int_V   F_{\alpha\beta}
F^{\alpha\beta} \sqrt{-g}\; d^4x,
\label{eq:Maxwell_action}
\end{equation}
where $F_{\alpha\beta}$ is the antisymmetric field-strength tensor
related to the four-vector potential, $A_{\mu}$, by
\begin{equation}
F_{\alpha\beta} = \nabla_\alpha A_\beta -  \nabla_\beta A_\alpha\, .
\end{equation} 
Here $\nabla$ represents covariant differentiation.

Varying the Maxwell action with respect to the vector potential and
setting the variation equal to zero leads to the source free inhomogeneous 
Maxwell equation for the electromagnetic field in curved spacetime,
\begin{equation}
\nabla^{\alpha}F_{\alpha\beta} = 0.
\label{eq:max_inhomo}
\end{equation}
Due to the Bianchi identities, the electromagnetic field also satisfies
the subsidiary condition,
\begin{equation}
\nabla_{[\alpha}F_{\beta\gamma]} = 0\, ,
\label{eq:max_homo}
\end{equation}
which is the homogeneous Maxwell equation. The combined set of equations
represents classical electrodynamics written in covariant form.  If we
insert the four-vector potential into both expressions, it is found that
the homogeneous Maxwell equation~(\ref{eq:max_homo}) is trivially satisfied. 
The inhomogeneous equation (\ref{eq:max_inhomo}) yields the second order
wave equation
\begin{equation}
\nabla^\alpha \nabla_\alpha A_\beta - \nabla_\beta \left(\nabla^\alpha A_\alpha
\right)- {R_\beta}^\alpha A_\alpha = 0 .
\label{eq:wave_eq}
\end{equation}
Here $R_{\alpha\beta}$ is the Ricci tensor which arises due to the
commutation relation for the covariant derivatives acting on a vector
field.

The stress-tensor for the classical electromagnetic field is found by
varying the Maxwell action with respect to the spacetime metric.  A
straightforward calculation yields
\begin{equation}
T_{\mu\nu}^{Maxwell} = F_{\mu\rho}{F_\nu}^\rho -{1\over 4} g_{\mu\nu}
F_{\alpha\beta}F^{\alpha\beta}\, .
\label{eq:Maxwell_stresstensor}
\end{equation} 

The field-strength tensor, the Maxwell equations and the stress tensor
are invariant under the gauge freedom
\begin{equation}
A_\alpha^{new} = A_\alpha^{old} - \nabla_\alpha \Lambda,
\end{equation}
where $\Lambda = \Lambda(x)$ is an arbitrary scalar function.  In classical 
electromagnetism, the correct choice of gauge can often simplify finding
the solution to the field equations.  In many cases, it is convenient
to choose the Lorentz gauge condition
\begin{equation}
\nabla^\alpha A_\alpha^{new} = 0,
\end{equation}
which immediately removes the middle term in the wave equation 
(\ref{eq:wave_eq}).
This can always be achieved by choosing $\Lambda$ to satisfy
\begin{equation}
\nabla^\alpha \nabla_\alpha \Lambda = \nabla^\alpha A_\alpha^{old}.
\end{equation}
It should be noted that there is still a restricted gauge freedom remaining
in that we can still add to the vector potential any function that satisfies
the homogeneous equation
\begin{equation}
\nabla^\alpha \nabla_\alpha \Lambda^{Hom.} = 0.
\end{equation}
As we shall see below, this restricted gauge freedom will be used to
impose the Coulomb gauge.  It should be noted that we will 
drop the identifiers of $new$ and $old$ in all further calculations

There is some difficulty is directly quantizing electrodynamics in the
form so far described.  If one does not specify a gauge, then any
four-vector wave equation like (\ref{eq:wave_eq}) will in general
have four orthonormal solutions (polarization states), $A_\alpha(\lambda;x)$
where $\lambda =$ 0, 1, 2 or 3.  In Minkowski spacetime the $\lambda = 0$
solution is typically the scalar photon polarization, $\lambda = 1$ and $2$
are the two transverse photon polarizations, and $\lambda = 3$ is the axial
photon polarization.  In curved spacetime the ``perfect'' separation of the 
modes
into these three ``distinct'' types is not always possible, but we will continue
to use the flat space nomenclature.  It is found that for one of the 
polarizations,
say $A_\alpha(0;x)$, there does not exist a conjugate momenta when the 
Hamiltonian is calculated.  This is a long known problem in flat spacetime
electrodynamics and there are several known approaches  which have been
developed to quantize the electromagnetic field that can be generalized
to curved spacetime.  The simplest is to work in a specific gauge 
\cite{Dimo92,Ford85}. A more elegant possibility is to use the Gupta-Bleuler
\cite{Gupta50,Gupta,Crisp_etal} formalism of indefinite metrics on the
Hilbert space of states.  Both of these forms of quantization are
discussed below.

\subsection{Direct quantization in the Coulomb gauge}

This is probably the simplest and most direct method of quantizing the 
electromagnetic field.  The problem so far stems from the fact that
the vector potential has four polarization states, while it is known
that the photons of the free field theory only come in two different
polarizations.  Thus, before the theory is quantized we would like to
remove the two superfluous polarizations at the classical level.  To do
this we require that solutions to the wave equation~(\ref{eq:wave_eq})
also satisfy the Lorentz gauge condition
\begin{equation}
\nabla^\alpha A_\alpha = 0.
\end{equation}
This removes one degree of freedom between the components of the vector
potential.  The next condition that we would like to require is that the
time component of the four-vector potential vanish in some frame.  To
accomplish this we let $\xi^\alpha$ be a timelike vector field.  Then we
require that $A_\alpha$ satisfy the additional condition
\begin{equation}
\xi^\alpha A_\alpha = 0.
\end{equation}
It is this second condition that can be ensured by the homogeneous part of
the gauge freedom. Also, note that is is not true that the Coulomb gauge
is noncovariant as is sometimes stated. 

In flat spacetime there is no preferred choice of $\xi^\alpha$, however
for the static metric of the form (\ref{eq:metric}), a natural choice is to
let $\xi^\alpha$ to be the global timelike Killing vector field.  This will
be the same Killing vector that will be used to define the positive frequency
mode functions.  Since $\xi^\alpha \propto (1,0,0,0)$, the net effect is
to set the $A_0$-component of the stress tensor equal to zero.  This solves
two problems simultaneously.  First it removes $A_0$ from the action, thus
there is no longer a problem of it not having a conjugate momenta.  Secondly,
it has reduced the physical degrees of freedom of the solution to the two
physically realizable photon states.

Canonical quantization is now straightforward.  The metric (\ref{eq:metric})
possesses a timelike killing vector, which allows us to write the the positive
frequency mode function solutions of the wave equation~(\ref{eq:wave_eq}) as
\begin{equation}
 A_\alpha (\lambda,{\bf k} ;{\bf x},t) = U_{\alpha} (\lambda,{\bf k}
;{\bf x})\; e^{-i\omega t},
\end{equation}
where ${\bf k}$ is the mode label for the propagation vector, $\lambda$
is the polarization state and $\omega = \omega({\bf k})$.  The four-vector
functions, $U_{\mu} (\lambda,{\bf k};{\bf x})$, are the spatial portion
of the solution of the wave equation and carry all the information about
the curvature of the spacetime.  In addition they satisfy 
\begin{equation}
\nabla^\alpha U_{\alpha} (\lambda,{\bf k};{\bf x}) = 0 = \nabla^\alpha
 \overline{U_{\alpha} (\lambda,{\bf k};{\bf x})}.
\end{equation}
The mode functions for the vector potential are normalized such that 
\begin{equation}
\left( A(\lambda,{\bf k}), A(\lambda',{\bf k}')\right) =
-i\int_{\Sigma} d\Sigma_\mu \left[ \overline{A_\nu (\lambda,\bf{k})} 
\,F^{\mu\nu}
(\lambda',{\bf k}') - \overline{F^{\mu\nu} (\lambda,\bf{k})}\,
A_\nu (\lambda',\bf{k}')\right] = \delta^{\lambda\lambda'}\delta^3({\bf k} -
{\bf k}'),
\label{eq:Coulomb_norm}
\end{equation}
where $d\Sigma_\mu = d\sigma \, n_\mu$ is a three-volume element in the 
Cauchy surface $\Sigma$ with unit normal $n^\mu$, thus each
mode contributes ${1\over 2}\omega$ to the vacuum expectation value of the
stress tensor before renormalization. The general solution to the vector
potential can then be expanded as
\begin{equation}
A_\mu({\bf x},t) = \sum_{\lambda=1}^{2}\int d^3{\bf k} \left(  a_{\lambda} 
({\bf k})
\,A_{\mu} (\lambda,{\bf k};{\bf x},t)  + 
a_{\lambda}^\dagger({\bf k})\, \overline{A_{\mu} (\lambda,{\bf k};{\bf x},t)}
\right)\,.
\end{equation}
Here $a_{\lambda}^\dagger({\bf k})$ and $a_{\lambda}({\bf k})$ are the  
creation and annihilation operators for the photon which obey the
commutation relations
\begin{equation}
{\big [} a_{\lambda}({\bf k}),a_{\lambda'}({\bf k}'){\big ]} = 0 =
{\big [} a_{\lambda}^\dagger({\bf k}),a_{\lambda'}^\dagger({\bf k}'){\big ]}
\end{equation}
and
\begin{equation} 
{\big [} a_{\lambda}({\bf k}),a_{\lambda'}^\dagger({\bf k}'){\big ]}=
\delta_{\lambda\lambda'} \delta({\bf k} - {\bf k}').
\end{equation}

The Fock representation of the number states can now be constructed from
the vacuum state denoted by $|0;0\rangle$ where the first slot is for
particles of polarization type 1 and the second slot is for polarization
type 2.  The vacuum state has the property
\begin{equation}
a_{\lambda}({\bf k}) | 0;0 \rangle = 0, \qquad\forall \; \{\lambda ,\,{\bf 
k}\}.
\end{equation}
One-particle states are obtained by acting on the vacuum with the creation
operator,
\begin{equation}
| 1_{\bf k};0\rangle = a^\dagger_1({\bf k}) | 0;0 \rangle \qquad
\mbox{and}\qquad
| 0;1_{\bf k}\rangle = a^\dagger_2({\bf k}) | 0;0 \rangle.
\end{equation}
Multi-particle states can likewise be created by repeated application of
the creation operators,
\begin{equation}
| ^1m_{{\bf k}_1}, \,\dots,\,^jm_{{\bf k}_j} ; 
  ^1n_{{\bf k}_1}, \,\dots,\,^jn_{{\bf k}_j}
\rangle= 
{\left( a^\dagger_{1}({\bf k}_1) \right)^{^1m}\dots
\left( a^\dagger_{1}({\bf k}_j) \right)^{^jm}
\left( a^\dagger_{2}({\bf k}_1) \right)^{^1n}\dots
\left( a^\dagger_{2}({\bf k}_j) \right)^{^jn}
\over
\left( ^1m! \dots {^jm}!\,^1n! \dots {^jn}!\right)^{1/2}}
|0;0\rangle,
\end{equation} 
where the ${\bf k}_1 ,{\bf k}_2,\,\dots,\, {\bf k}_j$ are all distinct.
The above state contains ${^1m}+{^2m}+\dots+{^jm}+{^1n}+{^2n}+\dots+{^jn}$
total particles where ${^1m}$ of them are of momentum ${\bf k}_1$ and
polarization 1, ${^1n}$  are of momentum ${\bf k}_1$ and polarization 2,
etc.  Effectively, the general number states are a direct product of elements
from two different Hilbert spaces, one for each of the polarization states.
In order to reduce the index notation to a more manageable form, define the
two vectors
\begin{equation}
{\bf m} = \left(^1m_{{\bf k}_1}, \,\dots,\,^jm_{{\bf k}_j}\right)
\qquad\mbox{and}\qquad
{\bf n} = \left(^1n_{{\bf k}_1}, \,\dots,\,^jn_{{\bf k}_j}\right),
\end{equation}
then the states can be written more simply as $|{\bf m};{\bf n}\rangle$.
The most general state that can then be written as a linear
superposition of all the possible number states,
\begin{equation}
|\psi\rangle = \sum_{{\bf m},{\bf n}}
c ({\bf m},{\bf n}) \,|{\bf m};{\bf n}\rangle,
\end{equation}
where $c ({\bf m},{\bf n})$ are complex coefficients and the sum is
assumed to range over all the allowed vectors of ${\bf m}$ and ${\bf n}$.
For the state to be properly normalized, the $c ({\bf m},{\bf n})$'s
must satisfy
\begin{equation}
\sum_{{\bf m},{\bf n}} |c ({\bf m},{\bf n})|^2 = 1.
\end{equation}

%%%%%%%%%%%%%%%%%%%%%%%%%%%%%%%%%%%%%%%%
\subsection{Gupta-Bleuler Formalism}
A more elegant form of quantization is to use the Gupta-Bleuler method of
imposing an indefinite metric on the Hilbert space of allowable states. 
We begin by forming the Gupta action,
\begin{equation}
S^{Gupta} = S^{Maxwell}+S^{G.B.},
\end{equation}
where $S^{Maxwell}$ is the Maxwell action given by 
Eq.~(\ref{eq:Maxwell_action}),
and the gauge breaking action is given by
\begin{equation}
S^{G.B.} = -{1\over 2} \int_V \left(\nabla^\alpha A_\alpha\right)^2 
\sqrt{-g} \;d^4x.\label{eq:GaugeBreak_action}
\end{equation}
Variation of the new action with respect to $A^\alpha$ yields the
wave equation,
\begin{equation}
\nabla^\alpha F_{\alpha\beta} + \nabla_\beta (\nabla^\alpha A_\alpha) = 0,
\end{equation}
which can be rewritten in terms of $A_\alpha$ as
\begin{equation}
\nabla^\alpha \nabla_\alpha A_\beta - {R_\beta}^\alpha A_\alpha = 0.
\label{eq:gupta_WE}
\end{equation}
This would correspond to Maxwell's equations if the field also satisfied
the Lorentz gauge condition.

There are four possible solutions (polarizations)
to the above wave equation.  First, there are the two physical polarizations
which are labeled with $\lambda = 1$ or $2$.  These two polarizations satisfy
the wave equation~(\ref{eq:gupta_WE}) and the Lorentz condition,
\begin{equation}
\nabla^\alpha A_\alpha(\lambda,{\bf k};{\bf x},t) = 0 \qquad\mbox{ for }
\qquad \lambda =1,2.
\label{eq:GB_physical_Lorentz}
\end{equation}
Thus, these two polarizations correspond to the two standard solutions to
Maxwell's equations.  The remaining two unphysical polarizations, labeled
with $\lambda = 0$ or $3$, also satisfy the wave equation~(\ref{eq:gupta_WE}),
but not necessarily the Lorentz condition.  For ultra-static spacetimes,
where $|g_{00}|=1$, the most natural choice is to use the scalar photon
polarization,
\begin{equation}
A_\alpha(0,{\bf k};{\bf x},t) = {1\over \omega}\left(\partial_t,0,0,0\right)
\phi({\bf k};{\bf x},t),
\end{equation} 
and the longitudinal photon polarization,
\begin{equation}
A_\alpha(3,{\bf k};{\bf x},t) = {1\over \omega}\left(0,\partial_{x^1}
,\partial_{x^2},\partial_{x^3}\right)
\phi({\bf k};{\bf x},t),
\end{equation} 
where $\phi({\bf k};{\bf x},t)$ is the solution to the massless, minimally
couples scalar wave equation,
\begin{equation}
\nabla^\alpha \nabla_\alpha \,\phi({\bf k};{\bf x},t) = 0.
\end{equation}
In the more general case of a static spacetime, it is useful to choose
the two orthogonal modes which satisfy the condition,
\begin{equation}
A_\alpha(0,{\bf k};{\bf x},t) + A_\alpha(3,{\bf k};{\bf x},t)
={1\over\omega} \nabla_\alpha \,\phi({\bf k};{\bf x},t).
\end{equation}
In both cases, the resulting modes then satisfy
\begin{equation}
\nabla^\alpha A_\alpha(0,{\bf k};{\bf x},t) = 
-\nabla^\alpha A_\alpha(3,{\bf k};{\bf x},t)
\end{equation}
and
\begin{equation}
F_{\alpha\beta}(0,{\bf k};{\bf x},t) = 
-F_{\alpha\beta}(3,{\bf k};{\bf x},t),
\label{eq:field_identity}
\end{equation}
for every momenta $\bf k$.  

In addition, if we define the generalized conjugate momenta,
\begin{equation}
\Pi^{\mu\nu} \equiv -(F^{\mu\nu} + g^{\mu\nu} \nabla^\alpha A_\alpha),
\end{equation}
then the modes are required to be orthogonal and normalized by
\begin{equation}
\left( A(\lambda,{\bf k}), A(\lambda',{\bf k}')\right) =
i\int_{\Sigma} d\Sigma_\mu \left[ \overline{A_\nu (\lambda,{\bf k})} 
\,\Pi^{\mu\nu}
(\lambda',{\bf k}') - \overline{\Pi^{\mu\nu} (\lambda,{\bf k})}\,
A_\nu (\lambda',{\bf k}')\right] = \eta^{\lambda\lambda'}\delta^3({\bf k} -
{\bf k}'),
\label{eq:Gupta_Normalize}
\end{equation}
where $d\Sigma_\mu = d\sigma \, n_\mu$ is a three-volume element  in the 
Cauchy surface $\Sigma$ with unit normal $n^\mu$,  and 
$\eta^{\lambda\lambda'} = \eta_{\lambda\lambda'}= $ diag(-1,1,1,1).
The general solution to $A_\mu$ then has the Fourier mode-decomposition
\begin{equation}
A_\mu({\bf x},t) = \sum_{\lambda=0}^{3}\int d^3{\bf k} \left(  a_{\lambda} 
({\bf k})
\,A_{\mu} (\lambda,{\bf k};{\bf x},t)  + 
a_{\lambda}^\dagger({\bf k})\, \overline{A_{\mu} (\lambda,{\bf k};{\bf x},t)}
\right)\,.
\end{equation}

If we wish to canonically quantize the field $A_\mu$, we impose the
equal-time commutation relations
\begin{equation}
\left[ A_\mu({\bf x},t), A_\nu({\bf x}',t)\right] = 0 =
\left[ \Pi^{t\mu}({\bf x},t), \Pi^{t\nu}({\bf x}',t)\right]
\end{equation}
and
\begin{equation}
\left[ A_\mu({\bf x},t), \Pi^{t\nu}({\bf x}',t)\right] = {i {\delta_\mu}^\nu
\over\sqrt{-g}}\delta^3({\bf x} - {\bf x}').
\end{equation}
Using the mode decomposition and the normalization condition, we find that
the above equal-time commutation relations are equivalent to
\begin{equation}
\left[ a_{\lambda} ({\bf k}), a_{\lambda'}^\dagger ({\bf k}') \right]
= \eta_{\lambda\lambda'}\delta^3({\bf k} - {\bf k}'),
\end{equation}
with all other commutators vanishing.

The state structure is similar in form to that found for the Coulomb gauge,
except there are now a greater number of allowable states due to the two
unphysical unphysical polarizations.  We now define the vacuum state as
$|0;0;0;0\rangle$ where the first slot is for photons of the unphysical polarization
$\lambda = 0$,  the second and third slots are for the two real photon
polarizations, and the final slot is for the unphysical polarization with
$\lambda = 3$.  The vacuum state vanishes if any of the four destruction 
operators act on it, and multi-particle states are again obtained by the
repeated application of the creation operators.  Unlike, the states for 
the Coulomb gauge quantization, the states of the Gupta-Bleuler formalism
have indefinite norm,
\begin{equation}    
\langle{{\bf l},{\bf m},{\bf n},{\bf p}}|{{\bf l}',{\bf m}',{\bf n}',{\bf p}'}\rangle
=(-1)^{^1l+^2l+\dots+^jl}\delta_{\bf l\, l'}\delta_{\bf m\, m'}\delta_{\bf n\, n'}
\delta_{\bf p\, p'},
\end{equation}
where we have added two new vectors, $\bf l$ and $\bf p$, for the 
unphysical photon polarization states. The most general state in the
Gupta-Bleuler formulation can be written as
superposition of all the particle number states as
\begin{equation} 
|\phi \rangle = \sum_{{\bf l},{\bf m},{\bf n},{\bf p}} c(
{{\bf l},{\bf m},{\bf n},{\bf p}})\, |{{\bf l},{\bf m},{\bf n},{\bf p}}\rangle.
\label{eq:superposition}
\end{equation}

In order for the Gupta-Bleuler formalism to be equivalent to Maxwell's theory,
we need to impose an additional condition on the Hilbert space of states; that
the expectation value of the Lorentz condition be satisfied for all physically
realizable states $|\phi\rangle$,
\begin{equation}
\langle \phi |\nabla^\alpha A_\alpha({\bf x},t) |\phi \rangle = 0.
\end{equation}
This condition can be accomplished simply by requiring that the states obey
\begin{equation}
\nabla^\alpha A^+_\alpha({\bf x},t) |\phi \rangle = 0,
\label{eq:GB_subsidiary}
\end{equation}
where $A^+_\alpha$ is the positive frequency part of $A_\alpha$.  The 
application of this condition to the state $|\phi \rangle$ above means
that the $c({{\bf l},{\bf m},{\bf n},{\bf p}})$'s with the same
total number of $\lambda = 0$ and $3$ photons of the same momenta are
related to one and other by
\begin{equation}
\sqrt{l_{\bf k}}\nabla^\alpha A_\alpha(0,{\bf k})
c\left( l_{\bf k},{\bf m},{\bf n},(p-1)_{\bf k}\right)
+\sqrt{p_{\bf k}}\nabla^\alpha A_\alpha(3,{\bf k})
c\left( (l-1)_{\bf k},{\bf m},{\bf n},p_{\bf k}\right)=0.
\end{equation}
Under this constraint the Hilbert space structure of the state $|\phi\rangle$
takes the form,
\begin{eqnarray}
|\phi\rangle &=& \cdots + c(0_{\bf k},{\bf m},{\bf n},0_{\bf k})
|0_{\bf k},{\bf m},{\bf n},0_{\bf k}\rangle +\cdots\nonumber\\
&&+ {c(1_{\bf k},{\bf m},{\bf n},0_{\bf k})\over \nabla^\alpha A_\alpha(3,{\bf 
k})}
\left[ \nabla^\alpha A_\alpha(3,{\bf k})|1_{\bf k},{\bf m},{\bf n},0_{\bf 
k}\rangle
-\nabla^\alpha A_\alpha(0,{\bf k})|0_{\bf k},{\bf m},{\bf n},1_{\bf k}\rangle
\right]+\cdots\nonumber\\
&&- {c(1_{\bf k},{\bf m},{\bf n},1_{\bf k})\over \sqrt{2}\nabla^\alpha 
A_\alpha(0,
{\bf k})\nabla^\alpha A_\alpha(3,{\bf k})}
\left[ \left(\nabla^\alpha A_\alpha(3,{\bf k})\right)^2|2_{\bf k},{\bf m},{\bf 
n},
0_{\bf k}\rangle-\sqrt{2}\nabla^\alpha A_\alpha(0,
{\bf k})\nabla^\alpha A_\alpha(3,{\bf k})|1_{\bf k},{\bf m},{\bf n},1_{\bf 
k}\rangle
\right.\nonumber\\
&&\qquad\left.-\left(\nabla^\alpha A_\alpha(0,{\bf k})\right)^2
|0_{\bf k},{\bf m},{\bf n},2_{\bf k}\rangle
\right]+\cdots\, .
\end{eqnarray}
With the definition of a new operator
\begin{equation}
O^\dagger({\bf k}) = \nabla^\alpha A_\alpha(3,{\bf k})\,a^\dagger_0({\bf k}) -
\nabla^\alpha A_\alpha(0,{\bf k})\,a^\dagger_3({\bf k}),
\end{equation}
it is possible to define a new set of states, $|{\bf m},{\bf n},{\bf 
q}\rangle$,
where each ${^iq}_{{\bf k}_i}$ in $\bf q$ is the total number of $\lambda = 0$ 
and
$3$ photons of momenta ${\bf k}_i$.  The new states are formed by the repeated
action of the operator $O^\dagger({\bf k})$ acting on the state with zero
unphysical photons,
\begin{equation}
|{\bf m},{\bf n},\{{^1q}_{{\bf k}_1},{^2q}_{{\bf k}_2},\dots, {^iq}_{{\bf 
k}_i}\}\rangle =
{\left(O^\dagger({{\bf k}_1})\right)^{^1q} \left(O^\dagger({{\bf 
k}_2})\right)^{^2q}
\dots\left(O^\dagger({{\bf k}_i})\right)^{^iq} \over( ^1q!\, ^2q!\dots\, 
^iq!)^{1/2}} 
|{\bf 0},{\bf m},{\bf n},{\bf 0}\rangle.
\end{equation}
The inner product of the new states are
\begin{equation}
\langle{\bf m},{\bf n},{\bf 0}|{\bf m},{\bf n},{\bf 0}\rangle =1
\end{equation}
for the states with no unphysical photons and
\begin{equation}
\langle{\bf m},{\bf n},q_{\bf k}|{\bf m},{\bf n},q_{\bf k}\rangle = 
\left\{ |\nabla^\alpha A_\alpha(0,{\bf k})|^2 - |\nabla^\alpha A_\alpha
(3,{\bf k})|^2 \right\}^q = 0
\end{equation}
for all other states.
It is now possible to rewrite the superposition of particle number
states~(\ref{eq:superposition}) with the embodiment of the 
supplementary condition~(\ref{eq:GB_subsidiary}) built in as
\begin{equation}
|\phi\rangle = \sum_{{\bf m},{\bf n},{\bf q}} b(
{{\bf m},{\bf n},{\bf q}})\, |{{\bf m},{\bf n},{\bf q}}\rangle\, .
\end{equation}

The stress tensor found from the Gupta action is
\begin{equation}
T^{Gupta}_{\rho\sigma}  = T^{Maxwell}_{\rho\sigma}+T^{G.B.}_{\rho\sigma},
\end{equation}
where $T^{Maxwell}_{\rho\sigma}$ is given by 
Eq.~(\ref{eq:Maxwell_stresstensor})
and the contribution to the stress-tensor from the gauge breaking term is
\begin{equation}
T^{G.B.}_{\rho\sigma} = - A_\rho\left(\nabla_\sigma \nabla^\alpha A_\alpha
\right) - A_\sigma \left( \nabla_\rho \nabla^\alpha A_\alpha\right) +
g_{\rho\sigma} \left[ A_\beta\nabla^\beta \nabla^\alpha A_\alpha +
{1\over 2} \left( \nabla^\alpha A_\alpha\right)^2\right].
\label{eq:GB_stresstensor}
\end{equation}
Due to the physical photon polarizations modes satisfying the Lorentz
condition~(\ref{eq:GB_physical_Lorentz}) and the Hilbert space of 
states satisfying the the subsidiary condition~(\ref{eq:GB_subsidiary}),
it is relatively straightforward to show that the expectation value of
the normal ordered gauge-breaking portion of the stress-tensor vanishes,
\begin{equation}
\langle\phi | : T^{G.B.}_{\rho\sigma} : |\phi \rangle = 0.
\end{equation}
In addition, due to the relationships between the $c({{\bf l},{\bf m},
{\bf n},{\bf p}})$ coefficients and Eq.~(\ref{eq:field_identity}), 
it is simple to show for the normal-ordered Maxwell portion of the
stress-tensor that the unphysical photon modes do not contribute, thus
\begin{equation}
\langle\phi| : T^{Maxwell}_{\rho\sigma} : | \phi \rangle =
\langle\psi| : T^{Maxwell}_{\rho\sigma} : | \psi \rangle,
\end{equation}
where
\begin{equation}
| \psi \rangle = \sum_{{\bf m},{\bf n}} c({0,{\bf m},{\bf n},0})\, 
|{0,{\bf m},{\bf n},0}\rangle = \sum_{{\bf m},{\bf n}} c({{\bf m},{\bf n}})\, 
|{{\bf m},{\bf n}}\rangle\, .
\end{equation}
Thus, the only physically observable states are the two physical photon
polarization states.   In summary we have
\begin{eqnarray}
\langle\phi | : T^{Gupta}_{\rho\sigma} : |\phi \rangle =
\langle\phi | : T^{Maxwell}_{\rho\sigma}+T^{G.B.}_{\rho\sigma} : |\phi\rangle
= \langle\psi| : T^{Maxwell}_{\rho\sigma} : | \psi \rangle.
\end{eqnarray}

%%%%%%%%%%%%%%%%%%%%%%%%%%%%%%%%%%%%%%%%%%%%%%%%%%%%%%%%%%%%%%%%%%%%%%%%%%%%%%%%%%
\goodbreak\section{Positivity Result}\label{sec:positivity}

In this section we prove the following inequality:   Let $M^{ij}$ be a real,
symmetric $n \times n$ matrix with non-negative eigenvalues. Further let
$P_i(\lambda,{\bf k})$ be a complex $n$-vector, which is a function of the
mode labels $\bf k$ and $\lambda$, Also, let $f(t)$ be a smooth,
non-negative function on ${\bf R}$ which decays rapidly at infinity, with
pointwise square root $f^{1/2}(t) = \sqrt{f(t)}$ and Fourier transform given by
\begin{equation}
\hat f(\omega) \equiv \int_{-\infty}^\infty dt \,f(t) \,e^{-i\omega t}.
\end{equation}
Then in an arbitrary quantum state  $|\psi \rangle$, the following inequality
holds,
\begin{eqnarray}
{\rm Re}\sum_{\lambda,\lambda'}\int d^3{\bf k} \; d^3{\bf k}' \,
\left\{ \hat f[\omega({\bf k}')-\omega({\bf k})] \langle a_\lambda^\dagger
({\bf k}) a_{\lambda'}({\bf k}') \rangle \overline{{P_i}(\lambda,{\bf k})}
M^{ij} P_j(\lambda',{\bf k}') \nonumber\right.\pm\\
\qquad\qquad\pm\hat f[\omega({\bf k})+\omega({\bf k}')]\left.\langle 
a_\lambda({\bf k}) a_{\lambda'}({\bf k}') \rangle P_i(\lambda,{\bf k})
M^{ij} P_j(\lambda',{\bf k}')\right\}
\nonumber\\
\qquad
\geq -{1\over 2\pi}\int_0^\infty d\nu \sum_{\lambda} \int d^3{\bf k}\;
\left| \widehat {f^{1/2}}[\nu+\omega({\bf k})]\right|^2 \overline{{P_i}
(\lambda,{\bf k})} M^{ij} P_j(\lambda,{\bf k}).
\label{eq:lowbound}
\end{eqnarray}

The above inequality is a generalization of the scalar
field positivity lemma derived by Fewster and colleagues 
\cite{Fe&E98,Fe&T99}.  In order to prove this relation,
first define the vector operator
\begin{equation}
\left[ Q^{\pm}_\nu\right]_i  =  \sum_{\lambda} \int d^3{\bf k}\,
\left\{\overline{g[\nu-\omega({\bf k})]} a_\lambda({\bf k}) 
P_i(\lambda,{\bf k})
\pm \overline{g[\nu+\omega({\bf k})]} a_\lambda^\dagger({\bf k}) 
\overline{{P_i}
(\lambda,{\bf k})} \right\},
\end{equation}
where
\begin{equation}
g(\omega) = {1\over\sqrt{2\pi}} \widehat{f^{1/2}}(\omega).
\end{equation}
From the definition of the convolution
\begin{equation}
\left( h_1 \star h_2\right)(\omega) = \int_{-\infty}^\infty d\omega'
\,h_1(\omega-\omega') \,h_2(\omega'),
\end{equation}
it follows that $(g\star g) = \hat f$.

Next, note that
\begin{equation}
M^{ij} = \sum_{\alpha =1}^n \kappa^\alpha V^i_{(\alpha)}\,V^j_{(\alpha)}\,,
\end{equation}
where the $V^i_{(\alpha)}$ are the eigenvectors of $M^{ij}$, and the
$\kappa^\alpha \geq 0$ are the corresponding eigenvalues. Now 
\begin{equation}
{\left\langle \left[ Q^{\pm}_\nu\right]_i^\dagger M^{ij} 
\left[ Q^{\pm}_\nu\right]_j \right\rangle =
\sum_{\alpha =1}^n \kappa^\alpha 
\left\langle \left[ Q^{\pm}_\nu\right]_i^\dagger  
V^i_{(\alpha)}\,V^j_{(\alpha)}
\left[ Q^{\pm}_\nu\right]_j \right\rangle = \sum_{\alpha =1}^n 
\kappa^\alpha 
|V^i_{(\alpha)}\, \left[ Q^{\pm}_\nu\right]_i |\psi\rangle |^2
\geq 0 \,.\label{eq:Lower_bound}}
\end{equation}
Furthermore, using the commutation relations and symmetrising the integrand
in $(\lambda,{\bf k})$ and $(\lambda',{\bf k}')$, we find
\begin{equation}
\int_0^\infty d\nu \langle \left[ Q^{\pm}_\nu\right]_i^\dagger M^{ij} 
\left[ Q^{\pm}_\nu\right]_j \rangle = \langle S^{\pm} \rangle +
\int_0^\infty d\nu \sum_{\lambda} \int d^3{\bf k} \;
\left|g[\nu+\omega({\bf k})]\right|^2 \,\overline{{P_i}(\lambda,{\bf 
k})}\,M^{ij}
P_i(\lambda,{\bf k}),
\label{eq:integral_exp}
\end{equation}
where
\begin{eqnarray}
\langle S^{\pm} \rangle&=& {\rm Re} 
\sum_{\lambda, \lambda'} \int d^3{\bf k} \; d^3{\bf k}' \, \left[
F({\bf k},{\bf k}') \langle a_\lambda^\dagger
({\bf k}) a_{\lambda'}({\bf k}') \rangle \,\overline{{P_i}(\lambda,{\bf k})}
M^{ij} P_j(\lambda',{\bf k}') \right.\nonumber\\
&&\qquad\qquad\pm \left.G({\bf k},{\bf k}')
\langle a_\lambda ({\bf k}) a_{\lambda'}({\bf k}') \rangle\,
{P_i}(\lambda,{\bf k}) M^{ij} P_j(\lambda',{\bf k}') \right]
\end{eqnarray}
and the functions $F$ and $G$ are given by
\begin{equation}
F({\bf k},{\bf k}') = \int_0^\infty d\nu\left\{ g[\nu-\omega({\bf k})]\,
\overline{g[\nu-\omega({\bf k}')]}
+\overline{g[\nu+\omega({\bf k})]} \,g[\nu+\omega({\bf k}'))\right\}
\label{eq:define_F}
\end{equation}
and
\begin{equation}
G({\bf k},{\bf k}') = \int_0^\infty d\nu\left\{ 
g[\nu+\omega({\bf k}))] \,\overline{g[\nu-\omega({\bf k}')]}
+\overline{g[\nu-\omega({\bf k})]} \,g[\nu+\omega({\bf k}')]\right\}.
\label{eq:define_G}
\end{equation}
The expressions for $F$ and $G$ may be simplified to 
\begin{eqnarray}
F({\bf k},{\bf k}') &=& \int_{-\infty}^\infty d\nu\,g[\omega({\bf 
k}')-\nu]
\,g[\nu-\omega({\bf k})]\nonumber\\
&=& (g\star g)[\omega({\bf k}')-\omega({\bf k})]\nonumber\\
&=& \hat f[\omega({\bf k}')-\omega({\bf k})]
\end{eqnarray}
and
\begin{equation}
G({\bf k},{\bf k}') = \hat f[\omega({\bf k})+\omega({\bf k}')].
\end{equation}
From Eq.(\ref{eq:Lower_bound}) we know that the right hand side of 
Eq.(\ref{eq:integral_exp}) is manifestly positive, so we conclude
that $\langle S^{\pm} \rangle$ obeys the following bound
\begin{eqnarray}
\langle S^{\pm} \rangle &\geq& - \int_0^\infty d\nu \sum_{\lambda} 
\int d^3{\bf k} \; \left|g[\nu+\omega({\bf k})]\right|^2\,\overline{{P_i}
(\lambda,{\bf k})}\, M^{ij} P_i(\lambda,{\bf k})\, , \nonumber\\ 
&=& -{1\over 2\pi}\int_0^\infty d\nu \sum_{\lambda} \int d^3{\bf k} \;
\left|\widehat{f^{1/2}}[\nu+\omega({\bf k})]\right|^2\,\overline{{P_i}
({\bf k},\lambda)}\,M^{ij}\, P_i(\lambda,{\bf k})\, ,
\end{eqnarray}
thus proving Eq.~(\ref{eq:lowbound}).

%%%%%%%%%%%%%%%%%%%%%%%%%%%%%%%%%%%%%%%%%%%%%%%%%%%%%%%%%%%%%%%%%%%%%%%%%%%%%%%%%%
\section{The Quantum Inequality}\label{sec:the_QI}

Consider a stationary observer whose four-velocity is given by
\begin{equation}
u^\mu = ( |g_{00}|^{-1/2}, 0, 0, 0 ).
\end{equation}
In both the simple quantization scheme using the Coulomb gauge and in the
Gupta-Bleuler quantization scheme, the energy density measured by
this observer is given by the Maxwell portion of the stress-tensor,
\begin{eqnarray}
\rho &=& T^{Maxwell}_{\mu\nu}u^\mu u^\nu\nonumber\\
&=& {1\over 2} \left[ {1\over |g_{00}|} F_{i0}\, g^{ij} \,F_{j0} + 
{1\over2}F_{ij}\,g^{il}\,g^{jm}\,F_{lm} \right].
\end{eqnarray}
Now make the identification
\begin{equation}
E_i = F_{i0} \qquad\mbox{and}\qquad B_i = {1\over 2}\varrho_{ijk} F_{jk}\, ,
\label{eq:field_def}
\end{equation}
where $\varrho_{ijk}$ is the completely antisymmetric Levi-Civita symbol.
The energy density can then be written as
\begin{equation}
\rho = {1\over 2}|g|^{-1/2}\left[ E_i\, \hat\epsilon^{ij} E_j +
B_i \left( \hat\epsilon^{ij} \right)^{-1} B_j \right]\, ,
\end{equation}
where $\hat\epsilon$ is an ordinary $3\times 3$ matrix with elements
\begin{equation}
\hat\epsilon = \hat\epsilon({\bf x}) = {\sqrt{-g}\over |g_{00}|}\left( 
  \begin{array}{lll}
         g^{11} & g^{12} & g^{13} \\
         g^{21} & g^{22} & g^{23} \\
         g^{31} & g^{32} & g^{33}
  \end{array}\right)\, .
\end{equation}
The definitions of $E_i$, $B_i$ and $\hat\epsilon$ have been shown
to recast the curved space Maxwell field equations into the form of
the Maxwell equations inside an anisotropic material medium in
Cartesian coordinates.  In this interpretation, $\hat\epsilon$ plays
the role of the dielectric tensor in the constitutive relations. 
We will not push this interpretation any further and refer the reader
to references \cite{Pleb60,Volk71,Mash73} for further discussion.

Upon substitution of the mode function expansion into the stress-tensor,
and making use of constitutive relations and the commutation relations
we find
\begin{eqnarray}
\rho & = &  |g|^{-1/2} \left\{{\rm Re}\!\!\sum_{\lambda\lambda'}\int 
d^3{\bf k}\, d^3{\bf k}'\,
\left[ a_{\lambda}^\dagger({\bf k}) a_{\lambda'}({\bf k}')\:
\overline{E_i(\lambda,{\bf k};{\bf x})}\; \hat\epsilon^{ij} \;
E_j(\lambda',{\bf k}';{\bf x}) e^{i(\omega-\omega')t}\right.\right.
\nonumber\\
&&\qquad\qquad\left. + a_{\lambda}({\bf k}) a_{\lambda'}({\bf k}')\:
E_i(\lambda,{\bf k};{\bf x})\; \hat\epsilon^{ij} \;
E_j(\lambda',{\bf k}';{\bf x})e^{-i(\omega+\omega')t} \right] 
\nonumber\\
&& + {\rm Re}\!\!\sum_{\lambda\lambda'}\int 
d^3{\bf k}\, d^3{\bf k}'\,
\left[ a_{\lambda}^\dagger({\bf k}) a_{\lambda'}({\bf k}')\:
\overline{B_i(\lambda,{\bf k};{\bf x})} \;  (\hat\epsilon^{ij})^{-1} \;
B_j(\lambda',{\bf k}';{\bf x}) \, e^{i(\omega-\omega')t}
\right. \nonumber\\&&\left. \qquad \qquad 
+ a_{\lambda} ({\bf k})a_{\lambda'}({\bf k}')\:
B_i(\lambda,{\bf k};{\bf x}) \;  (\hat\epsilon^{ij})^{-1} \;
B_j(\lambda',{\bf k}';{\bf x}) \, \, e^{-i(\omega+\omega')t}
\right] \nonumber\\
&& + \left.{1\over 2} \sum_{\lambda}\int d^3{\bf k} \left[ 
E_i(\lambda,{\bf k};{\bf x})\; \hat\epsilon^{ij} \;
\overline{E_j(\lambda,{\bf k};{\bf x})} +B_i(\lambda,{\bf k};{\bf x})
\;  (\hat\epsilon^{ij})^{-1} \; \overline{B_j(\lambda,{\bf k};{\bf x})}\right]
\right\},
\label{eq:energy_density} 
\end{eqnarray}
where 
\begin{equation}
E_i(\lambda,{\bf k};{\bf x}) = \partial_i\, U_0(\lambda,{\bf k};{\bf x})
+ i\,\omega({\bf k})\, U_i(\lambda,{\bf k};{\bf x})
\end{equation}
and
\begin{equation}
B_i(\lambda,{\bf k};{\bf x}) = \varrho_{ijl}\, \partial_j \,
U_l(\lambda,{\bf k};{\bf x}).
\end{equation}
The last line of Eq.~(\ref{eq:energy_density}) is the vacuum
self-energy of the photons.  As was the case for the scalar field,
we will look at the difference between the energy in an arbitrary
state relative to the vacuum energy using the normal order prescription,
{\it i.e.},
\begin{equation}
:\rho: = \rho -  \rho_{\rm vacuum}.
\label{eq:normal_order}
\end{equation}
It is now our intention to show that given a temporal sampling function $f(t)$
then the sampled energy density defined by
\begin{equation}
\Delta\tilde\rho = \int_{-\infty}^{\infty} dt \;\langle : \rho ({\bf x},t)
 : \rangle f(t) \, ,
\label{eq:Samp_Edensity_definition}
\end{equation} 
is bounded from below.  Using Eqs.~(\ref{eq:energy_density}) and
(\ref{eq:normal_order}), along with the definitions of $F({\bf k},{\bf k}')$
and $G({\bf k},{\bf k}')$ given by Eqs.~(\ref{eq:define_F}) and
(\ref{eq:define_G}), the sampled energy density is
\begin{eqnarray}
\Delta\tilde\rho &=& {1\over |g_{00}|}\, {\rm 
Re}\!\!\sum_{\lambda\lambda'}\int 
d^3{\bf k}\, d^3{\bf k}'\, \omega({\bf k})\,\omega({\bf k}')\left[ F({\bf 
k},{\bf k}') 
\langle a_{\lambda}^\dagger({\bf k}) a_{\lambda'}({\bf k}')\rangle\:
\overline{E_i(\lambda,{\bf k};{\bf x})}\; g^{ij} \; E_j(\lambda',{\bf k}';{\bf 
x})\,
\right.\nonumber\\
&&\qquad\qquad\left.+ G({\bf k},{\bf k}') \langle a_{\lambda}({\bf k})
a_{\lambda'}({\bf k}')\rangle\: E_i(\lambda,{\bf k};{\bf x})\; g^{ij}
\; E_j(\lambda',{\bf k}';{\bf x})\, \right]\nonumber\\
&& + \left|{g_{00}\over g}\right| \, {\rm Re}\!\!\sum_{\lambda\lambda'}\int 
d^3{\bf k}\, d^3{\bf k}'\, \left[ F({\bf k},{\bf k}') \langle 
a_{\lambda}^\dagger({\bf k}) a_{\lambda'}({\bf k}')\rangle\:
\overline{B_i(\lambda,{\bf k};{\bf x})}\; (g^{ij})^{-1} \; B_j({\bf k}',
\lambda';{\bf x}) \right. \nonumber\\&&\left. \qquad \qquad 
+ G({\bf k},{\bf k}')\langle a_{\lambda} ({\bf k})a_{\lambda'}({\bf k}')
\rangle\: B_i(\lambda,{\bf k};{\bf x})\;(g^{ij})^{-1} \; B_j
(\lambda',{\bf k}';{\bf x})\right].
\end{eqnarray}
Clearly, both parts of the above expression are of the form $\langle S^\pm
\rangle$, so we may apply the bound (\ref{eq:lowbound}) with $M^{ij} = g^{ij}$
and $P_i(\lambda,{\bf k}) = E_i(\lambda,{\bf k};{\bf x})$ for the first
part of the expression and $M^{ij} = (g^{ij})^{-1}$
and $P_i(\lambda,{\bf k}) = B_i(\lambda,{\bf k};{\bf x})$ for the second
part of the expression.  This yields a difference inequality of
\begin{eqnarray}
\Delta\tilde\rho &\geq&  -{1\over 2\pi} \int_0^\infty d\nu
\sum_\lambda \int d^3{\bf k} \left| \widehat{f^{1/2}}[ \nu + \omega
({\bf k})] \right|^2 \left[{1\over |g_{00}|} 
\overline{E_i(\lambda,{\bf k};{\bf x})}\; g^{ij} \; E_j(\lambda,{\bf k};{\bf 
x})
\right.\nonumber\\
&&\qquad\qquad+\left.\left|{g_{00}\over g}\right| \overline{B_i(\lambda,{\bf 
k};
{\bf x})}\; (g^{ij})^{-1} \; B_j(\lambda,{\bf k};{\bf x})\right].
\label{eq:EM_Diff_Ineq}
\end{eqnarray}
This expression is similar in form to the mode function expansion
of the scalar field quantum inequality found by Fewster and colleagues
\cite{Fe&E98,Fe&T99}. The quantum inequality, Eq.~(\ref{eq:EM_Quant_Ineq}),
is found by adding the suitably renormalized vacuum energy density to
the above expression.

%{\it The QI also reduces to an averaged weak energy type integral in the
%infinite sampling time limit.}

%%%%%%%%%%%%%%%%%%%%%%%%%%%%%%%%%%%%%%%%%%%%%%%%%%%%%%%%%%%%%%%%%%%%%%%%%%%%%%%%%%
\section{Examples}\label{sec:examples}
%%%%%%%%%%%%%%%%%%%%%%%%%%%%%%%%%%%%%%%%
\subsection{Minkowski Spacetime}
This quantum inequality is easily evaluated in Minkowski spacetime with
no boundaries. Using quantization in the Coulomb gauge, the four-vector
mode functions are
\begin{equation}
A_\alpha (\lambda,{\bf k} ;{\bf x},t) = \left(0, {\bf A}(\lambda,{\bf k} ;
{\bf x},t)\right),
\end{equation}
where
\begin{equation}
{\bf A}(\lambda,{\bf k} ;{\bf x},t) = {i\over \sqrt{2\omega (2\pi)^3 }}\,
\hat{\varepsilon}_{\bf k}^\lambda \, e^{i( {\bf k \cdot x} - \omega t) },
\end{equation}
$\hat{\varepsilon}_{\bf k}^\lambda$ is a unit electric polarization
vector and $\omega = \sqrt{\bf k \cdot k}$.  Due to the Coulomb gauge 
condition,
the propagation vector is orthogonal to the polarization vector, {\it i.e.}
\begin{equation}
{\bf k} \cdot \hat{\varepsilon}_{\bf k}^\lambda = 0.
\end{equation}
A third, orthogonal unit vector along the magnetic field direction is
defined by
\begin{equation}
\hat{b}_{\bf k}^\lambda = \hat{\bf k} \times \hat{\varepsilon}_{\bf 
k}^\lambda. 
\end{equation}
Inserting the mode functions into Eq.~(\ref{eq:EM_Diff_Ineq}), and using
$g^{ij} = \delta^{ij}$, we find
\begin{eqnarray}
\tilde\rho &\geq& -{1\over (2\pi)^4} \int_0^\infty d\nu
\sum_\lambda\int d^3{\bf k} \left\{ \widehat{f^{1/2}}[ \nu + \omega
({\bf k})] \right\}^2 \,\omega ({\bf k})\,\left[\hat{\varepsilon}_{\bf 
k}^\lambda
\cdot\hat{\varepsilon}_{\bf k}^\lambda + \hat{b}_{\bf k}^\lambda\cdot
\hat{b}_{\bf k}^\lambda \right],\nonumber\\
&=& -{4\over (2\pi)^3} \int_0^\infty d\nu \int_0^\infty d\omega
\,\omega^3 \left\{ \widehat{f^{1/2}}[ \nu + \omega] \right\}^2,
\end{eqnarray}
where we have made a change of variable in the momentum integration
to spherical coordinates and have already carried out the
angular integration and summation over polarization
states.  The next step is to make another change of variable
\begin{equation}
u = \nu + \omega, \qquad v = \omega,
\label{eq:coord_change}
\end{equation}
to find
\begin{eqnarray}
\tilde\rho &\geq & -{4\over (2\pi)^3} \int_0^\infty du  \left[ 
\widehat{f^{1/2}}(u) 
\right]^2\int_0^u dv \;v^3,\nonumber\\
&=&-{1\over 2(2\pi)^3} \int_{-\infty}^\infty du  \left[ 
u^2\;\widehat{f^{1/2}}(u) 
\right]^2.
\end{eqnarray}
Using Parsaval's Identity, the quantum inequality is found to be
\begin{equation}
\tilde\rho\geq-{1\over 8\pi^2} \int_{-\infty}^\infty dt \left[ 
{d^2\over dt^2} f^{1/2}(t)\right]^2 .
\end{equation}
This is the most general expression for the quantum inequality in Minkowski
spacetime with an arbitrary sampling function.  For the choice of a Lorentzian
sampling function~(\ref{eq:lorentzian}) with characteristic width $t_0$, it
is straightforward to calculate
\begin{equation}
\tilde\rho \geq - {27\over 1024 \,\pi^2 \,t_0^4}.
\end{equation}
This is a slightly stronger result, by $9/64$, than the inequality proven
by Ford and Roman \cite{F&Ro97} using an alternative method.  Comparison
with the quantum inequality for the scalar field in Minkowski space derived
by Fewster and Eveson \cite{Fe&E98}, shows that the electromagnetic field
quantum inequality in Minkowski space always differs by a factor of two. 
This is a result of the electromagnetic field having two polarization
degrees of freedom, unlike the scalar field which has only one, and the both
the scalar and electromagnetic field modes having the same energy spectrum.
Electromagnetic field quantum inequalities for various sampling functions
are summarized in Table~1.

%%%%%%%%%%%%%%%%%%%%%%%%%%%%%%%%%%%%%%%%
\subsection{Rindler Spacetime}

Next, we would like to find the quantum inequality in Rindler spacetime.
We begin with the Minkowski space length element,
\begin{equation}
ds^2 = -dt^2+dx^2+dy^2+dx^2.
\end{equation}
Next we apply the coordinate transformation
\begin{eqnarray}
t&=& \xi \sinh \eta,\nonumber\\
x&=& \xi \cosh \eta,
\end{eqnarray}
to arrive at the Rindler length element
\begin{equation}
ds^2 = -\xi^2 d\eta^2 + d\xi^2+dy^2+dz^2.
\end{equation}
In this form, the metric is static, but the $g_{00}$ component is
not a constant, so we can not quantize the theory in the Coulomb
gauge but must use the Gupta-Bleuler formalism.  Thus we are looking
for mode solutions to the vector wave equation~(\ref{eq:gupta_WE}).
These, have been calculated by Candelas and Deutsch~\cite{Ca&DE77}
for the two physical polarizations.  The unphysical solutions have
also been calculated \cite{Hig_etal,Moretti}. The modes can be
conveniently expressed in terms of the mode solutions to the
massless scalar field wave equation in Rindler space,
\begin{equation}
\left(-{1\over \xi^2}\partial_\eta^2 +{1\over \xi}\partial_\xi\,
\xi \partial_\xi +\partial_y^2 +\partial_z^2\right) \phi(x) = 0.
\end{equation} 
The positive frequency scalar mode solutions, normalized for
the Klein-Gordon inner product of scalar fields, are \cite{Full73}
\begin{equation}
\phi(\omega,k_y,k_z;x)={2\over(2\pi)^2}(\sinh\omega\pi)^{1/2}
K_{i\omega}(\beta\xi)\,e^{i(k_y y+ k_z z - \omega\eta)},
\end{equation}
where $\beta = \left(k_y^2 + k_z^2\right)^{1/2}$ and $K_{i\nu}(x)$
are the modified Bessel functions of the second kind (Macdonald
functions) of {\it imaginary order}.  The two physical modes that
are important to our calculations are the transverse electric
modes (TE),
\begin{equation}
A_{\alpha}(1,\omega,k_y,k_z;x)= {1\over\beta}\left(
0,\,0,\,\partial_z,\,-\partial_y \right)\phi(\omega,k_y,k_z;x),
\end{equation}
and the transverse magnetic modes (TM),
\begin{equation}
A_{\alpha}(2,\omega,k_y,k_z;x)= {1\over\beta}\left(
\xi\partial_\xi,\,{1\over\xi}\partial_\eta,\,0,\,0
\right)\phi(\omega,k_y,k_z;x).
\end{equation}
These two modes are properly normalized with respect to 
Eq.~(\ref{eq:Gupta_Normalize}) and are also orthogonal.
If they are inserted into Eq.~(\ref{eq:EM_Diff_Ineq})
for the difference inequality, and after a little algebra, we find
\begin{eqnarray}
\Delta\tilde\rho &\geq&  -{1\over \pi} \int_0^\infty d\nu
\int_0^\infty d\omega \left| \widehat{f^{1/2}}[ \nu + \omega] \right|^2 
\int_{R^2}dk_y\,dk_z\, \left[ \beta^2 {\overline\phi}\phi +
{\omega^2\over\xi^2}{\overline\phi}\phi+
(\partial_\xi {\overline\phi})(\partial_\xi\phi)\right],\nonumber\\
&=& -{1\over \pi^4} \int_0^\infty d\nu
\int_0^\infty d\omega \left| \widehat{f^{1/2}}[ \nu + \omega] \right|^2 
\left({\omega^2\over\xi^2}+{1\over 4\xi}\partial_\xi\,\xi
\partial_\xi\right)\sinh(\pi\omega)
\int_0^\infty d\beta \,\beta\,K^2_{i\omega}(\beta\xi),
\end{eqnarray}
where we have switched to polar coordinates to carry out the angular
portion of the $dk_y\,dk_z$ integrals. With the aid of Eq.~6.521.3 of
\cite{Gradshteyn},
it is easily demonstrated that
\begin{equation}
\int_0^\infty d\beta \,\beta\,K^2_{i\omega}(\beta\xi) =
{\pi\omega \over 2 \xi^2 \sinh(\pi\omega)}.
\end{equation}
Thus
\begin{eqnarray}
\Delta\tilde\rho &\geq&  -{1\over 2\pi^3 \xi^4} \int_0^\infty d\nu
\int_0^\infty d\omega \,\omega(\omega^2+1)
\left| \widehat{f^{1/2}}[ \nu + \omega] \right|^2,\nonumber\\[8pt]
&=& -{1\over 16\pi^3 \xi^4}\left\{ \int_{-\infty}^\infty
\left| u^2 \widehat{f^{1/2}}(u) \right|^2 du + 2\int_{-\infty}^\infty
\left| u \widehat{f^{1/2}}(u) \right|^2 du\right\},\nonumber\\[8pt]
&=&-{1\over 8\pi^2 \xi^4}\left\{ \int_{-\infty}^\infty
\left[ {d^2\over d\eta^2} {f^{1/2}}(\eta) \right]^2 d\eta + 
2\int_{-\infty}^\infty \left[ {d\over d\eta}
{f^{1/2}}(\eta) \right]^2 d\eta\right\},
\label{eq:rindler_DI}
\end{eqnarray}
where we have again changed the variables of integration in accordance with
Eq.~(\ref{eq:coord_change}) in the second line and used Parsaval's Identity
to arrive at the third line.  The quantum inequality is found by adding
the Rindler space vacuum energy density \cite{Ca&DE77},
\begin{equation}
\rho_{\rm vacuum} = -{1\over\pi^2\xi^4}\int_0^\infty d\omega {\omega^3+\omega
\over e^{2\pi\omega}-1} = -{11\over 240\pi^2\,\xi^4},
\end{equation}
to the above expression.  For the Lorentzian sampling function,
Eq.~(\ref{eq:lorentzian}), and the definition of the proper time of the static
observer, $\tau = \xi\eta$, we find
\begin{equation}
\Delta\tilde\rho \geq -{27\over 1024 \pi^2 \tau_0^4}\left[ 1+ {32\over 27}
\left({\tau_0\over\xi}\right)^2\right].
\end{equation}
Once again we find that the Rindler space difference inequality for the 
electromagnetic field is twice that of the scalar field result found by
Fewster and Eveson \cite{Fe&E98} for the same reason as discussed in the
previous example.  The electromagnetic field quantum inequalities for
other sampling functions are also summarized in Table~1.

We need to be careful about the interpretation of this quantum inequality
in Rindler spacetime, as it appears that both the vacuum energy density
and the expression for the difference inequality, Eq.~(\ref{eq:rindler_DI}),
diverge in the limit as $\xi\rightarrow 0$.  This does not mean  the
quantum inequality fails on the particle horizon in Rindler spacetime. This
divergence is really a pathology of the coordinates and spacetime trajectory
used. Recall that the quantum inequality found above is for a static observer
in the Rindler coordinates.  This trajectory is not that of a geodesic observer
but one undergoing constant acceleration.  A ``static'' observer at $\xi = 0$
would require a constant infinite acceleration, an impossible scenario.
The divergence in the quantum inequality expresses this.
We can then ask what is the quantum inequality along the worldline of a
geodesic observer in Rindler space?  Well, a geodesic observer in Rindler
spacetime is the same as a constant velocity geodesic in Minkowski
spacetime, with the resulting quantum inequality in the geodesic
observer's rest frame already found in the preceding Minkowski space example.
It is obvious that there is nothing ``unique'' happening as the geodesic
observer crosses the point is space which is associated with the particle
horizon in Rindler coordinates.  Thus, in Rindler space, the quantum inequality 
along a geodesic worldline does not fail.

%%%%%%%%%%%%%%%%%%%%%%%%%%%%%%%%%%%%%%%%
\subsection{Static Einstein Spacetime}

Finally, we study the quantum inequality in the static closed universe where
the length element is given by
\begin{equation}
ds^2 = -dt^2 + a^2 \left[ d\chi^2 +\sin^2\chi\left(d\theta^2+\sin^2\theta
d\varphi^2\right)\right],
\end{equation}
and $a$ is the radius of the universe.  The modes of the electromagnetic
field in this spacetime have been studied by various authors 
\cite{Lifs46,Park72,Mash73}.  In terms of the vector potential, the mode
solutions are the vector spherical harmonics on $\rm{S}^3$ with harmonic
time dependence.  In a fashion similar to the previous examples, the
four-vector mode functions can be found from a scalar function that
satisfies the partial differential equation
\begin{equation}
\left(\nabla^\alpha \nabla_\alpha - {2\cos\chi\over a^2 \sin\chi}\partial_\chi
\right) \psi_{nlm}(t,\chi,\theta,\varphi) = 0,
\end{equation} 
which is not the scalar wave equation in the Einstein universe.  The scalar
mode solutions are
\begin{equation}
\psi_{nlm}(t,\chi,\theta,\varphi) = V_{nl}(\chi)\,Y_{lm}(\theta,\varphi)
\,e^{-i\omega_n t},
\end{equation}
where $\omega_n = n/a$ and $Y_{lm}(\theta,\varphi)$ are the scalar
spherical harmonics on $\rm{S}^2$. The functions $V_{nl}(\chi)$ are
defined as
\begin{equation}
V_{nl}(\chi) = {2^l l! \sqrt{(n-l-1)!} \over \sqrt{
l (l+1) \pi (n+l)!}}\,\sin^{l+1}\chi\, C^{l+1}_{n-l-1}(\cos\chi),
\end{equation}
where $C^\lambda_\eta(x)$ are the Gegenbauer polynomials
as defined in \cite{Gradshteyn}. The primary quantum number $n$ ranges
over the integers greater than one, i.e. $n=2,3,4,\dots$. For a given $n$
there are $n^2-1$ harmonic states with the same energy labeled by the 
quantum numbers, $l=1,\dots,n-1$ and $n= -l, -l+1, \dots, 0, \dots\, l-1, l$.

The two physical four-vector potential modes are the electric J-pole modes,
\begin{equation}
A_{\alpha}(1,n,l,m;x)= {1\over n}\left(0, {l(l+1)\over\sin^2\chi},
\partial_\chi \partial_\theta, \partial_\chi \partial_\varphi \right)
\psi_{nlm}(t,\chi,\theta,\varphi),
\end{equation}
and the magnetic J-pole modes,
\begin{equation}
A_{\alpha}(2,\omega,k_y,k_z;x)= \left(0, 0, {1\over\sin\chi} \partial_\varphi,
\sin\chi \partial_\theta\right) \psi_{nlm}(t,\chi,\theta,\varphi),
\end{equation}
both of which satisfy the vector wave equation~(\ref{eq:wave_eq}), the
Lorentz gauge and Coulomb gauge conditions, and are orthonormal.

Inserting these mode into Eq.~(\ref{eq:EM_Diff_Ineq}) yields
\begin{eqnarray}
\Delta\tilde\rho &\geq& -{1\over\pi} \int_0^\infty d\nu \sum_{n=2}^\infty
\left| \widehat{f^{1/2}}[ \nu + \omega_n] \right|^2 \sum_{l=1}^\infty
{1\over a^4 \sin^2\chi}  \left\{ l (l+1) 
\left[{1\over 2}\partial_\chi^2 + 2n^2\right] V_{nl}^2 
\sum_{m=-l}^l\overline{Y_{lm}}Y_{lm} \right. \nonumber\\
&&\left.\qquad+{1\over 2 } \left[ \left(\partial_\chi 
V_{nl} \right)^2 + n^2 V_{nl}^2\right]
\left({1\over\sin\theta} \partial_\theta \sin\theta \partial_\theta
+{1\over\sin^2\theta}\partial_\varphi^2\right)\sum_{m=-l}^l
\overline{Y_{lm}}Y_{lm}\right\}.
\end{eqnarray}
However, the spherical harmonics satisfy an addition theorem,
\begin{equation}
\sum_{m=-l}^l \overline{Y_{lm}}Y_{lm} = {2l+1\over 4\pi},
\end{equation}
which is independent of the $\theta$ and $\varphi$ coordinates, thus
the terms in the expression for the difference inequality involving 
derivatives with respect to $\theta$ and $\varphi$ will vanish.  The
remaining terms can then be written as
\begin{equation}
\Delta\tilde\rho \geq -{1\over 2 \pi^2 a^4} \int_0^\infty d\nu \sum_{n=2}^\infty
\left| \widehat{f^{1/2}}[ \nu + \omega_n] \right|^2 \left( n^2 +
{1\over 4 \sin^2\chi}\partial_\chi^2 \sin^2\chi\right)
\sum_{l=1}^{n-1} {(2l+1)l(l+1) \over \sin^2\chi} V_{nl}^2.
\end{equation}
The Gegenbauer polynomials also satisfy an addition theorem,
Eq.~8.934.4 of \cite{Gradshteyn}, which for our case can be written as
\begin{equation}
\sum_{l=0}^{n-1} {(2l+1) 2^{2l} (l!)^2 (n-l-1)!  \over (l+l)!}\left[
\sin^l\chi \, C^{l+1}_{n-l-1}(\cos\chi) \right]^2 = n.
\end{equation}
Using this in the difference inequality leads to
\begin{eqnarray}
\Delta\tilde\rho &\geq& -{1\over 2 \pi^3 a^4} \int_0^\infty d\nu \sum_{n=2}^\infty
\left| \widehat{f^{1/2}}[ \nu + \omega_n] \right|^2 \left( n^2 +
{1\over 4 \sin^2\chi}\partial_\chi^2 \sin^2\chi\right) \left[ n -{1\over n}
\left(\sin n\chi\over \sin\chi\right)^2\right],\nonumber\\
&=&-{1\over 2 \pi^3 a^3} \int_0^\infty d\nu \sum_{n=2}^\infty
\omega_n (n^2-1) \left| \widehat{f^{1/2}}[ \nu + \omega_n] \right|^2 .
\end{eqnarray}
The resulting expression is spatially invariant, as expected in a homogeneous
and isotropic universe.  In addition, it has the general form of a sum over
all the energies times the multiplicity for each energy times the Fourier 
transform of the square root of the sampling function, a form similar to that
found by Fewster and Teo \cite{Fe&T99} for the scalar field in both the
three-dimensional closed universe and in the four-dimensional static
Robertson-Walker spacetimes.  In order to find the quantum inequality, we
need to add to the above expression the renormalized vacuum energy density
for the electromagnetic field which is found to be \cite{Ford76}
\begin{equation}
\rho_{\rm vacuum} = {11\over 240 \pi^2 a^4}.
\end{equation}

\begin{figure}[bh]
\begin{center}
%\leavevmode\epsfxsize=5in
\epsfbox{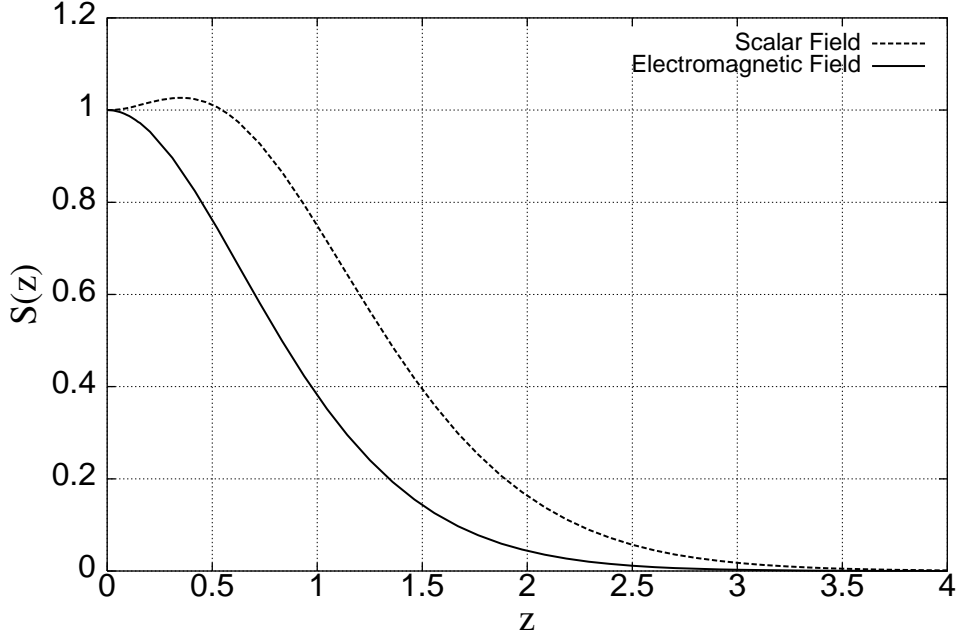}
\end{center}
\caption[]
{Plot of the scale functions for a Lorentzian sampling function in
the four-dimensional static Einstein universe.  The solid line is
the Electromagnetic field result, while the dotted line is the scalar
field result.  Note, for small $z=t_0/a$ both scale functions approach
one, while for large $z$ they decay to zero.}
\label{fig:closed_scale_function}
\end{figure}

When the difference inequality is evaluated for the Lorentzian sampling
function we find
\begin{equation}
\Delta\tilde\rho \geq - {27\over 1024 \pi^2 t_0^4} \, S_{EM}(t_0/a),
\end{equation}
where $S_{EM}(z)$ is the scale function for the closed universe given
by
\begin{equation}
S_{EM}(z) = {2048 \over 27 \pi^2} \; z^4 \sum_{n=2}^\infty n (n^2-1) \int_{nz}^\infty
K_0^2(u) \, du\, ,
\end{equation}
and $K_0(u)$ is the zero-order modified Bessel function of the second kind.
It is straightforward to evaluate this function numerically and is plotted
in Figure~\ref{fig:closed_scale_function}.  For sampling times very small
compared to the radius of the universe, the scale function is approximately
one, for which we effectively recover the flat space quantum inequality.
This makes sense because over such sampling times the region of the universe
over which the observer moves is indistinguishable from Minkowski space.
However, for sampling times on the order of, or larger than the radius of the
universe, the observer (and thus the quantum inequality) has time to ``see''
the large scale structure of the universe.  Thus the scale function changes
appreciably away from one.

It should also be pointed out that unlike the Minkowski and Rindler spacetime
examples, the quantum inequality for the electromagnetic field is not simply
twice that of the scalar field quantum inequality. In both of the previous cases,
the spacetimes are flat with the Riemann curvature term in the wave equation
vanishing.
Therefore, the electromagnetic wave equation~(\ref{eq:wave_eq}) in the Lorentz
gauge can be reduced to the scalar field wave equation. Thus, the energy spectra
are identical for the scalar and electromagnetic fields in each spacetime with
the factor of two coming from the degeneracy of electromagnetic field having two
orthogonal polarization states.  However, for the Einstein universe, and in
curved spacetimes in general, the energy spectrum for the scalar and
electromagnetic field modes are not the same,  thus the scalar and
electromagnetic quantum inequalities have different forms.

Using the work of Fewster and Teo \cite{Fe&T99}, the scalar field
difference inequality in the Einstein closed universe with a
Lorentzian sampling function is
\begin{equation}
\Delta\tilde\rho \geq - {27\over 2048 \pi^2 t_0^4} \, S_{scalar}(t_0/a),
\end{equation}
where
\begin{equation}
S_{scalar}(z) = {2048 \over 27 \pi^2} \; z^4 \sum_{n=0}^\infty \sqrt{n(n+2)}
(n+1)^2 \int_{\sqrt{n(n+2)}z}^\infty K_0^2(u) \, du.
\end{equation}
This scale function is also plotted in Figure~\ref{fig:closed_scale_function}
where we again see the generic behavior of the scale function being one for
small values of $t_0/a$ and decaying for large values.  However, unlike the
electromagnetic case which is monotonically decreasing function, the scalar
case has a bump which peaks at $t_0/a \sim 0.75$ and then smoothly
decays. The bump is due to the $n=1$ term in the summation, a term which has
no electromagnetic counterpart.  If this term is removed from the summation,
the remaining portion of the scale function does result in a monotonically 
decreasing behavior more akin to, but not exactly like the electromagnetic
case.  At present, it is not know if the bump in the scalar case has any
physical meaning, as no state has yet been demonstrated which actually
achieves this bound, although it may be a good guess that such a state would
include $n=1$ modes. There has also been an alternative conjecture that
the bump may be an artifact of the inequalities not being optimal. In
either case, further research on the scalar field quantum inequality
should eventually clarify this issue.

%%%%%%%%%%%%%%%%%%%%%%%%%%%%%%%%%%%%%%%%%%%%%%%%%%%%%%%%%%%%%%%%%%%%%%%%%%%%%%%%%%

\begin{center}{\bf Acknowledgments}\end{center}
I would like to thank C.J. Fewster, L.H. Ford, E. Poisson and T.A. Roman
for useful discussions and comments on the manuscript.  In addition, I
would like to thank George Leibbrandt for his aid on the quantization
section of the manuscript. This work was supported by the National
Sciences and Engineering Research Council of Canada.

\newpage
%%%%%%%%%%%%%%%%%%%%%%%%%%%%%%%%%%%%%%%%%%%%%%%%%%%%%%
\begin{center}
\renewcommand{\arraystretch}{2.5}
\noindent\begin{tabular}{|c|c|c|c|}
\hline\multicolumn{2}{|c|}{\bf Sampling Function} & {\bf Minkowski Spacetime}
 & {\bf Rindler Spacetime}\\ \hline
Lorentzian & $ t_0\over\pi(t^2+t_0^2) $ & $\tilde{\rho} \geq -{27\over 1024
\pi^2 t_0^4}$ &  $\;\;\Delta\tilde{\rho} \geq -{27\over 1024\pi^2 \tau_0^4} \left[
1+{32\over 27}\left({\tau_0\over\xi}\right)^2\right]$\\[8pt] \hline
Lorentzian$^2$ & $2 t_0^3 \over\pi(t^2+t_0^2)^2$ & $\tilde{\rho} \geq -{3\over 16
\pi^2 t_0^4}$ &  $\Delta\tilde{\rho} \geq -{3\over 16\pi^2 \tau_0^4} \left[
1+{2\over 3}\left({\tau_0\over\xi}\right)^2\right]$\\[8pt] \hline
Gaussian & ${1\over\sqrt{\pi}t_0} e^{-(t/t_0)^2}$ & $\tilde{\rho} \geq -{3\over 32
\pi^2 t_0^4}$ &  $\Delta\tilde{\rho} \geq -{3\over 32\pi^2 \tau_0^4} \left[
1+{4\over 3}\left({\tau_0\over\xi}\right)^2\right]$ \\[8pt] \hline
Cosine$^4$    &\renewcommand{\arraystretch}{1} 
$\left\{\begin{array}{ll}{4\over 3t_0}
\cos^4\left( {\pi t\over 2 t_0}\right)&-t_0<t<t_0\\
0&\mbox{elsewhere}
\end{array}\renewcommand{\arraystretch}{2.5}
\right.$ & $\tilde{\rho} \geq -{\pi^2\over 96 t_0^4}$ & 
$\Delta\tilde{\rho} \geq -{\pi^2\over 96 \tau_0^4} \left[
1+{8\over \pi^2}\left({\tau_0\over\xi}\right)^2\right]$\\[8pt] \hline
\end{tabular}
\end{center}
\vspace*{0.1in}
{\indent {TABLE 1.} Electromagnetic field quantum inequalities in Minkowski 
spacetime and difference inequalities in Rindler spacetime calculated for
various unit area sampling functions.}
\vfill\vfill
%%%%%%%%%%%%%%%%%%%%%%%%%%%%%%%%%%%%%%%%%%%%%%%%%%%%%%
\vfill
%%%%%%%%%%%%%%%%%%%%%%%%%%%%%%%%%%%%%%%%%%%%%%%%%%%%%%

\end{document}